\documentstyle[pra,aps,multicol,epsfig]{revtex}

\newcommand\be{\begin{eqnarray}}
\newcommand\ee{\end{eqnarray}}
\newcommand\ba{\begin{array}}
\newcommand\ea{\end{array}}
\def\r{\rangle}
\def\l{\langle}
\def\T{{\rm Tr}}
\def\cH{{\cal H}}

\def\cI{{\cal I}}

\def\cE{{\cal E}}

\def\cA{{\cal A}}
\def\openone{{\it I}}

\begin{document}
\title{Process reconstruction: From unphysical to physical maps via maximum likelihood}
\author{M\'ario Ziman$^{1,2}$, Martin Plesch$^{1}$, Vladim\'\i r Bu\v zek$^{1,2}$, and Peter \v{S}telmachovi\v{c}$^{1}$
}
\address{
${}^{1}$
Research Center for Quantum Information,
Slovak Academy of Sciences,
D\'ubravsk\'a cesta 9, 845 11 Bratislava, Slovakia \\
${}^{2}$ Faculty of Informatics, Masaryk University, Botanick\'a 68a,
602 00 Brno, Czech Republic
}
\date{14 January 2005}

\maketitle

\begin{abstract}
We show that the method of maximum likelihood (MML) provides us with an efficient scheme for reconstruction of
quantum channels from incomplete measurement data.
By construction this scheme always results in estimations of channels that are completely positive.
Using this property we use the MML for a derivation of physical approximations of un-physical
operations. In particular, we analyze the optimal approximation of the universal NOT gate as well as a physical approximation
of a quantum nonlinear polarization rotation.

\end{abstract}

%%%%%%%%%%%%%%%%%%%%%%%%%%%%%%%%%%%%%%%%%%%%%%%%%%%%%%%%%%%%%%%%%%%%%%%%%%%%%
\begin{multicols}{2}
%%%%%%%%%%%%%%%%%%%%%%%%%%%%%%%%%%%%%%%%%%%%%%%%%%%%%%%%%%%%%%%%%%%%%%%%%%%%

Any quantum dynamics \cite{nielsen,presskill}, i.e. the process
that is described by a completely positive (CP) map of a
quantum-mechanical system can be probed in two different ways.
Either we use a single entangled state of a bi-partite system
\cite{dariano}, or we use a collection of linearly independent
single-particle test states \cite{poyatos,nielsen} (that form a basis
of the vector space of all hermitian operators). Given the
fragility of entangled states in this Letter we will focus our
attention on the process reconstruction using only single-particle
states.

The task of a process reconstruction is to determine
an unknown quantum channel (a ``black box'') using
correlations between known input states and results of measurements
performed on these states that have been transformed by the channel.
The linearity of quantum dynamics implies
that the channel $\cE$ is exhaustively described by its action
 $\varrho_j\to\varrho_j^\prime=\cE[\varrho_j]$ on
a set of basis states, i.e. a collection of linearly independent
states $\varrho_j$,
that play a role of {\it test states}.
Therefore, to perform a reconstruction of the channel
${\cal E}$ we have
to perform a complete state tomography \cite{nielsen} of $\varrho_j^\prime$.
The number of test states equals $d^2$, where $d=\dim\cH$ is the dimension
of the Hilbert space associated with the system. Consequently, in order to reconstruct a channel
we have to determine $d^2(d^2-1)$ real
parameters, i.e. 12 numbers in the case of qubit ($d=2$).

In what follows we will assume that
test states can be prepared on demand perfectly. Nevertheless, the reconstruction of the channel ${\cal E}$ can
be affected by the lack of required information  due to the following reasons:
{\it i)} each test state is represented by a finite ensemble,
correspondingly, measurements
performed at the output can result in  an approximate estimation of transformed test states;
 {\it ii)} the set of test states is not complete;
and {\it iii)} incomplete measurements on transformed test states are performed.
In these cases some of the parameters that determine the map ${\cal E}$ cannot be
deduced perfectly from the measured data. In order to accomplish the channel
reconstruction additional criteria have to be considered.

In this Letter we will pay attention to the case {\it i)},
that is typical for experiments - one
cannot prepare an infinite ensemble, so
frequencies of the measured outcomes are only approximations of
probability distributions. Consequently, the reconstruction of
output states  $\varrho_j^\prime$ can lead us to unphysical
conclusions about the action of the quantum channel.
As a result we can find a negative operator $\varrho_j^\prime$,
or a channel $\cE$,
which is not CP.

In what follows we will introduce and compare
two schemes how to perform a channel reconstruction with insufficient measurement data.
Firstly, we will consider a rather straightforward ``regularization'' of the reconstructed
unphysical map. Secondly we will exploit the method of
{\it maximum likelihood} (MML) to perform an estimation of the channel. We will use
these methods to perform a reconstruction of maps based on numerical simulation of
the anti-unitary {\it universal NOT operation} (U-NOT) \cite{werner}.
This is a linear, but  not a CP map and we will show how our regularization methods
will result in optimal physical approximations of the U-NOT operation. We will conclude that the MML
is a tool that provide us with approximations of non-physical operations. In order to demonstrate the power of this approach
we will also apply it
to obtain an approximation of a nonlinear quantum-mechanical map, the so called
{\it nonlinear polarization rotation} (NPR) \cite{vinegoni}.

%%%%%%%%%%%%%%%%%%%%%%%%%%%%%%%%%%%%%%%%%%%%%%%%%%%%%%%%%%%%%%%%%%%%%%%%%%%%%
{\em Structure of qubit channels.}
%%%%%%%%%%%%%%%%%%%%%%%%%%%%%%%%%%%%%%%%%%%%%%%%%%%%%%%%%%%%%%%%%%%%%%%%%%%%%
Quantum channels are described by
linear trace-preserving CP maps $\cE$ defined on a
set of  density operators \cite{nielsen,presskill,ruskai}. The complete
positivity is guaranteed
if the operator $\Omega_\cE=\cE\otimes\cI[P_+]$ is a valid quantum state
[$P_+$ is a projection onto a maximally entangled state].
Any qubit channel $\cE$ can be imagined as an
affine transformation of
the three-dimensional Bloch vector $\vec{r}$
 (representing a qubit state),
i.e. $\vec{r}\to\vec{r}^\prime= T\vec{r}+\vec{t}$, where
$T$ is a real 3x3 matrix and $\vec{t}$ is a translation \cite{ruskai}.
This form guarantees that the
transformation $\cE$ is hermitian and trace preserving. The
CP condition defines (nontrivial) constraints on possible
values of involved parameters. In fact, the set of all CP
trace-preserving maps forms a specific convex subset of all affine
transformations.

The matrix $T$ can be written in the so-called
singular-value decomposition, i.e.
$T=R_UDR_V$ with $R_U,R_V$ corresponding to orthogonal rotations and
$D={\rm diag}\{\lambda_1,\lambda_2,\lambda_3\}$ being diagonal
where $\lambda_k$ are the singular values of $T$. This means that any
map $\cE$ is a member of less-parametric family of maps
of the ``diagonal form'' $\Phi_\cE$, i.e.
$\cE[\varrho]=U\Phi_\cE[V\varrho V^\dagger]U^\dagger$ where
$U,V$ are unitary operators.
The reduction of parameters is very helpful,
and most of the properties (including complete positivity)
of $\cE$ is reflected
by the properties of $\Phi_\cE$. The map $\cE$ is CP only if
$\Phi_\cE$ is. Let us note that $\Phi_\cE$ is determined not only by
the matrix $D$, but also by a new translation vector $\vec{\tau}=R_U\vec{t}$,
i.e. under the action of the map $\Phi_\cE$ the
Bloch sphere transforms as follows $r_j\to r_j^\prime=\lambda_j r_j+\tau_j$.

\begin{figure}
\begin{center}
\includegraphics[width=3.5cm]{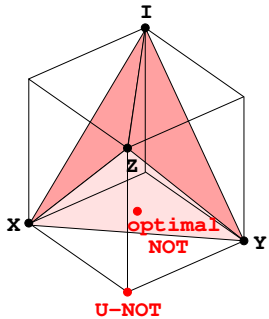}
\smallskip
\caption{Unital CP maps
are embeded in the set of all positive unital maps (cube). The CP maps form
a tetrahedron with four unitary transformations in its corners
(extremal points) $I,x,y,z$  corresponding to the Pauli $\sigma$-matrices.
The un-physical
U-NOT operation ($\lambda_1=\lambda_2=\lambda_3=-1$) and its optimal completely positive
approximation quantum universal NOT gate ($\lambda_1=\lambda_2=\lambda_3=-1/3$) are shown.}
\label{unital_cp}
\end{center}
\end{figure}

A special class of CP maps are the unital maps,
that transform the total mixture into itself. In this case
$\vec{t}=\vec{\tau}=\vec{0}$,
and the corresponding
map $\Phi_\cE$ is uniquely specified
by just three real parameters. The positivity of the transformation $\Phi_\cE$
results into conditions $|\lambda_k|\le 1$. On the other hand, in order to fulfill the
 CP condition
we need that the
four inequalities $|\lambda_1\pm\lambda_2|\le|1\pm\lambda_3|$ are satisfied.
These conditions specify a tetrahedron lying inside a cube
of all positive unital maps. In this case the extreme points represent four unitary
transformations $\openone,\sigma_x,\sigma_y,\sigma_z$
(see Fig.\ref{unital_cp}).
%%%%%%%%%%%%%%%%%%%%%%%%%%%%%%%%%%%%%%%%%%%%%%%%%%%%%%%%%%%%%%%
%%%%%%%%%%%%%%%%%%%%%%%%%%%%%%%%%%%%%%%%%%%%%%%%%%%%%%%%%%%%%%%
{\em Qubit channel estimation and its regularization.}
%%%%%%%%%%%%%%%%%%%%%%%%%%%%%%%%%%%%%%%%%%%%%%%%%%%%%%%%%%%%%%%
Reconstructions of states and processes share
many  features. Therefore we briefly remind us basic concept of the state reconstruction
using  {\it finite}
ensembles of identically prepared states. In this case one can obtain from estimated mean values
of observable a {\it negative} density operator of a qubit $\varrho=\frac{1}{2}(\openone+\vec{r}\cdot\vec{\sigma})$
with $|\vec{r}|>1$. The reconstructed
operator has always unit trace, but the associated vector $\vec{r}$
can point {\em out} of the Bloch sphere. One can argue that the proper physical state
is the closest one to the reconstructed operator, i.e. a pure state with $\vec{r}_c$
pointing into the same direction. Formally it corresponds
to a multiplication $\vec{r}$ by some constant $k$, i.e. $\vec{r}_c=k\vec{r}$.
The correction by $k$ can be expressed as
\be
\varrho_c = k\varrho+(1-k)\frac{1}{2}\openone =
\frac{1}{2}(\openone+k\vec{r}\cdot\vec{\sigma}) \,
\ee
and it can be understood as
a convex addition of the total mixture $\frac{1}{2}\openone$
represented by the center of the Bloch sphere, i.e.
$\vec{0}=(0,0,0)$. In other words, the correction consists of the addition of
completely random and equally distributed events (clicks) to
outcome statistics, i.e. {\it addition of random noise}.

As we have seen from above, an important role in the
state reconstruction is played by the total mixture $\frac{1}{2}\openone$, which is
an average over all possible states.
An average over all CP maps is the map $\cA$ that transforms the whole state space into
the total mixture, i.e. $\cA[\varrho]=\frac{1}{2}\openone$ \cite{ziman}.

The reconstruction of qubit channels consists of the known preparation
of (at least) four linearly independent test states $\varrho_j$
and a state reconstruction of corresponding four output states $\varrho_j^\prime$.
The process estimation based on the correlations $\varrho_j\to\varrho_j^\prime$
is certainly trace-preserving
(if all $\varrho_j^\prime$ are positive). Though the complete positivity
is problematic. The average channel $\cA$ can be used
to correct (``regularize'') improper estimations $\cE$ to obtain a CP qubit channel $\cE_c$
\be
\label{3}
\cE_c=k\cE+(1-k)\cA=\left(
\ba{cc}
1 & \vec{0} \\
k\vec{t} & kT
\ea
\right)\, .
\ee
This method of channel regularization uses the same principle
as the method for states, i.e. it is associated with an addition
of random noise into  data.

Let us try to estimate what is the critical value of
$k$, i.e. the amount of noise that surely corrects any positive map.
Trivially, it is enough to set $k=0$. In this case we completely
ignore the measured data and the corrected map is $\cA$.
However, we are interested in some nontrivial lower bound, i.e. in the largest
possible value of $k$ that guarantees the complete positivity.
Let us consider, for simplicity, that the map under consideration is  unital.
Then the worst case of a positive map that is not CP,
is represented by the universal NOT operation.

%%%%%%%%%%%%%%%%%%%%%%%%%%%%%%%%%%%%%%%%%%%%%%%%%%%%%%%%%%%%%%%%%%%%%%%%%%%%%
{\em Universal NOT.}
%%%%%%%%%%%%%%%%%%%%%%%%%%%%%%%%%%%%%%%%%%%%%%%%%%%%%%%%%%%%%%%%%%%%%%%%%%%%%
The logical NOT operation can be generalized into the quantum domain as
a unitary transformation $|0\r\to|1\r,|1\r\to|0\r$. However, this
map is basis dependent and does not transform all qubit states $|\psi\r$
into their (unique) orthogonal complements $|\psi_\perp\r$.
Such {\it universal NOT} ($\cE_{\tt NOT}:|\psi\r\to|\psi_\perp\r$)
is associated with the inversion of the Bloch sphere, i.e.
$\vec{r}\to -\vec{r}$, which is not a CP map. It represents an unphysical
transformation specified by $\lambda_1=\lambda_2=\lambda_3=-1$.
The distance (see Fig.\ref{unital_cp}) between this
map and the tetrahedron of completely positive maps is extremal,
i.e. it is most un-physical map
among linear transformations and can
be performed only approximatively.
A quantum ``machine'' that optimally
implements an approximation of the universal
NOT has been introduced in Ref.~\cite{werner}.
The machine is represented by a map
$\tilde{\cE}_{\tt NOT}={\rm diag}\{1,-1/3,-1/3,-1/3\}$.
The distance \cite{ziman} between the U-NOT and its optimal physical approximation reads
\be
\label{distance}
d(\tilde{\cE}_{\tt NOT},\cE_{\tt NOT})=\int_{states}
{\rm d}\varrho\T |(\cE_{\tt NOT}-\tilde{\cE}_{\tt NOT})[\varrho]|=1/3\, .
\ee

The CP conditions
imply that the minimal amount of noise necessary for a regularization of the universal NOT gate
corresponds to the value of
$k=1/3$, i.e. $\lambda_1=\lambda_2=\lambda_3=-1/3$
(see Fig.\ref{unital_cp}). The channel representing this point
corresponds to the best CP approximation of the
universal NOT operation, i.e. to the {\it optimal universal NOT machine}
originally introduced in Ref.~\cite{werner}.

One way how to interpret
the ``regularization'' noise is to assume that the qubit channel is influenced  by other quantum systems (the physics behind a dilation theorem
\cite{werner}).

The reason why we have to consider a noise in a reconstruction of quantum maps is that we deal with incomplete measurement
statistics (e.g., test states are represented by finite ensembles).
As a result, the reconstructed assignment
$\varrho_j\to\varrho_{j}^\prime={\cal{E}}[\varrho_{j}]$
is determined not only by the properties
of the map $\cal{E}$ itself but also by the character
of the estimation procedure. In this situation, the map itself can be
unphysical, but if we request that the estimation procedure is such that the complete positivity of the estimated map
is guaranteed then the result of the estimation is a physical approximation of an unphysical operation.
In order to proceed we assume the method of maximum likelihood.

{\it The maximum likelihood
 method} is a general estimation scheme \cite{fisher,hradil}
that has already been considered
for a reconstruction of quantum operations from incomplete data.
It has been studied
by Hradil and Fiur\'a\v sek
 \cite{fiurasek}, and by Sachci  \cite{sachci} (criticized in Ref.
\cite{hradil_kritizuje}).
The task of the
maximum likelihood in the process reconstruction is to find out
a map $\cE$, for which the {\it likelihood} is maximal.
By the definition we assume that
the estimated map has to be CP.  Let us now briefly describe the principal
idea in more details.

Given the measured data represented by couples $\varrho_k,F_k$
($\varrho_k$ is one of the test states and $F_k$ is a positive operator
corresponding to the outcome of the measurement used in the $k$th run of
the experiment) the likelihood functional is defined by the formula
\be
L(\cE)=-\log\prod_{k=1}^N p(k|k)=
-\sum_{k=1}^N \log\T \cE[\varrho_k]F_{k}\; ,
\label{5}
\ee
where $N$ is the total number of ``clicks'' and
we used $p(j|k)=\T\cE[\varrho_k]F_j$ for a conditional
probability of using test state $\varrho_k$ and observe the outcome $F_j$.
The aim is to find a physical map $\cE_{est}$
that maximizes this function, i.e. $L(\cE_{est})=\max_\cE L(\cE)$.
This variational task is usually performed  numerically.

%%%%%%%%%%%%%%%%%%%%%%%%%%%%%%%%%%%%%%%%%%%%%%%%%%%%%%%%%%%%%%%
\begin{figure}
\begin{center}
\includegraphics[width=5.5cm]{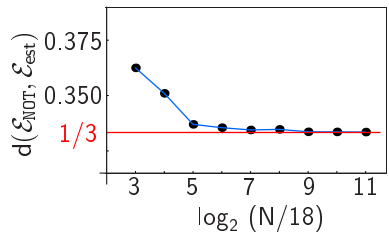}
\smallskip
\caption{The distance $d(\cE_{\tt NOT},\cE_{est})$
as a function of the number of measured outcomes $N$ in the logarithmic scale.
We used 6 input states
(eigenvectors of $\sigma_x,\sigma_y,\sigma_z$)
and we ``measured'' operators $\sigma_x,\sigma_y,\sigma_z$. The distance converges
to the theoretical value $1/3$ that corresponds to the optimal
universal NOT.
}
\label{reconstruction}
\end{center}
\end{figure}
%%%%%%%%%%%%%%%%%%%%%%%%%%%%%%%%%%%%%%%%%%%%%%%%%%%%%%%%%%%%%%%

%%%%%%%%%%%%%%%%%%%%%%%%%%%%%%%%%%%%%%%%%%%%%%%%%%%%%%%%%%%%%%%
\begin{figure}
\begin{center}
\includegraphics[width=6.5cm]{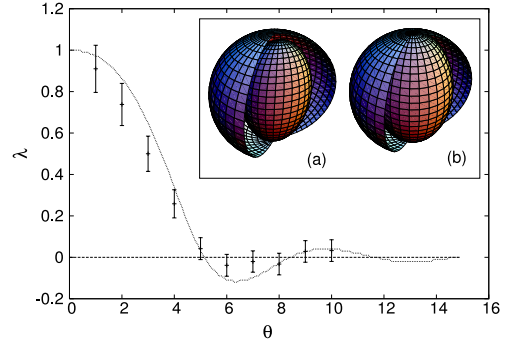}
\caption{We present analytical as well as numerical results
of an approximation of a non-linear map ${\cal E}_{\theta}$ given by Eq.~(\ref{eq6}) for different values
of the parameter $\theta$. The numerical (``experimental'')
results shown in the
graph in terms of a set of discrete points with error bars
are obtained via the MML. For every
point ($\theta$), the nonlinear operation was applied to 1800 input states
that have been chosen randomly (via a Monte Carlo method).
These input states have been transformed
according to the nonlinear transformation (\ref{eq6}). Subsequently
simulations of random projective measurements have been performed.
With these ``experimental'' data a linear operation was numerically
searched for, which maximizes the probability to obtain the same
results. The resulting approximation specified by a value of $\lambda$
(error bars shown in the graph represent the variance
in outcomes for subsequent runs) transforms the
original Bloch sphere as it is shown in an inset
 for a particular value $\theta=3$. The figure (a) corresponds
to result obtained by MML, and the figure (b) has been obtained
via analytic calculations.
We see that the original Bloch sphere is
transformed into an ellipsoid, one axis of which is significantly
longer than the remaining two
axes, that are of a comparable length. The mean of these two
lengths corresponds to the parameter $\lambda$ that specifies the map.
The analytical approximation $\tilde{\cal E}_{\theta}$
of the nonlinear NPR map is characterized by the parameter
$\lambda$ that is plotted  (solid line) in the figure
as a function of the parameter $\theta$.
}
 \label{fig3}
\end{center}
\end{figure}
%%%%%%%%%%%%%%%%%%%%%%%%%%%%%%%%%%%%%%%%%%%%%%%%%%%%%%%%%%%%%%%%%%%%%%%%%%%%%%%%%%%%%%%%%%%%%%%%%%%%%%%%%%%%%%%%%%%%

%%%%%%%%%%%%%%%%%%%%%%%%%%%%%%%%%%%%%%%%%%%%%%%%%%%%%%%%%%%%%%%%%%%%%%%%%%%%%
{\em Numerical results.}
%%%%%%%%%%%%%%%%%%%%%%%%%%%%%%%%%%%%%%%%%%%%%%%%%%%%%%%%%%%%%%%%%%%%%%%%%%%%%
Our approach is different from those described in Refs.~\cite{fiurasek,sachci,hradil_kritizuje}
in the way,
how we find the maximum of the functional defined in Eq.(\ref{5}).
The parametrization of $\cE$ itself, as defined by Eq.(\ref{3}), guarantees
the trace-preserving condition. The CP condition
is introduced as an {\em external} boundary for a Nelder-Mead simplex
scheme. Therefore, there is no need to introduce Lagrange multipliers.
We use the eigenstates of $\sigma_x,\sigma_y,\sigma_z$ as the
collection of test states. The data are generated as
(random) results of three projective measurements
$\sigma_x,\sigma_y,\sigma_z$ applied in order to perform the output state
reconstruction.
In order to analyze the convergence of the method
we have performed
the reconstruction for different number of clicks and compare
the distance between the original map
$\cE_{\tt NOT}$ and the estimated map $\cE_{est}$.
The result is plotted in Fig.\ref{reconstruction}, where we can see that
the distance converges to $1/3$ as calculated in Eq.(\ref{distance}).
For $N=100\times 18$ clicks, i.e.  each measurement
is performed 100 times per particular input state,
the algorithm leads us to the map
\be
\cE_{est}=\left(
\ba{cccc}
1 & 0 & 0 & 0 \\
-0.0002 & -0.3316 & -0.0074 & 0.0203 \\
 0.0138 & -0.0031 & -0.3334 & 0.0488 \\
-0.0137 &  0.0298 & -0.0117 & -0.3336
\ea
\right)
\ee
which is very close [$d(\cE_{est},\cE_{app})=0.0065$]
to the best approximation of the NOT operation, i.e.
$\cE_{app}={\rm diag}\{1,-1/3,-1/3,-1/3\}$.

We conclude that for large $N$ the MML reconstruction gives us the same result as a theoretical prediction derived in
Ref.~ \cite{werner}.
>From here it follows that the MML helps us not only to
estimate the map when just incomplete data are available, but also serves as a tool to
derive physical approximations of unphysical maps.
The reconstruction procedure guarantees
that the estimation/approximation is physical.
In order to illustrate the power of this approach we will find
an approximation of a non-linear quantum mechanical
transformation that is even ``more'' unphysical
than the linear though antiunitary U-NOT operation.

%%%%%%%%%%%%%%%%%%%%%%%%%%%%%%%%%%%%%%%%%%%%%%%%%%%%%%%%%%%%%%%
{\it Nonlinear polarization rotation }
%%%%%%%%%%%%%%%%%%%%%%%%%%%%%%%%%%%%%%%%%%%%%%%%%%%%%%%%%%%%%%%
Let us consider a nonlinear transformation of a qubit
defined by the relation \cite{vinegoni}:
\be
\cE_{\theta}[\varrho]=e^{i\frac{\theta}{2}\l\sigma_z\r_\varrho\sigma_z}
\varrho e^{-i\frac{\theta}{2}\l\sigma_z\r_\varrho\sigma_z}\,
\label{eq6}.
\ee
Unlike the universal NOT this map is nonlinear.
Four test states are not sufficient to get the whole
information about the action of nonlinear maps.
Consequently, the
fabricated data must use all possible input states
 (that cover the whole Bloch sphere) as test states,
but still we use only three
different measurements performed on outcomes that are sufficient for the state
reconstruction. We note that
a straightforward regularization via the addition of noise
cannot result in a CP map unless the
original map is not completely suppressed by the noise, i.e. the regularization
leads to a trivial result $\cE=\cA$. However, as we shall see,
the maximum likelihood approach gives us a reasonable and
nontrivial approximation of the transformation (\ref{eq6}).

Firstly, we present an analytic derivation of a physical approximation
of $\cE_\theta$. This approximation is the closest physical map
$\tilde{\cE}_\theta$,
i.e. $d(\tilde{\cE}_\theta,\cE_\theta)=\min$.
The map $\cE_\theta$ exhibits two  symmetries:
the continuous U(1) symmetry (rotations around the $z$-axis)
and the discrete $\sigma_x$ symmetry (rotation around the $x$-axis by $\pi$).
The physical approximation $\tilde{\cE}_\theta$ should possess
these properties as well. Exploiting these symmetries the possible
transformations of the Bloch vector are restricted as follows
$x\to\lambda x,y\to\lambda y,z\to pz$.
In the process of minimalization the parameter $p$ behaves trivially
and equals to one. It means that $\tilde{\cE}_\theta$ is of the form
$\cE_\lambda={\rm diag}\{1,\lambda,\lambda,1\}$.
Our task is to minimize the distance $d(\cE_\theta,{\cE}_\lambda)=\int d\varrho
|\cE_\theta[\varrho]-\cE_\lambda[\varrho]|$ and to find the physical
approximation $\tilde{\cE}_\theta$,
i.e. the functional dependence of $\lambda$ on $\theta$.

We plot the parameter $\lambda$ that specifies the best physical
approximation of the NPR map in Fig.~\ref{fig3}.
In the same figure we also present a result of the maximum
likelihood estimation of the NPR map based on a finite number of ``measurements'' (for technical details see the figure captions).
We conclude that the MML
is in an excellent agreement with our analytical calculations.
%%%%%%%%%%%%%%%%%%%%%%%%%%%%%%%%%%%%%%%%%%%%%%%%%%%%%%%%%%%%%%%

%%%%%%%%%%%%%%%%%%%%%%%%%%%%%%%%%%%%%%%%%%%%%%%%%%%%%%%%%%%%%%%

In summary, we have shown that theoretical methods that  are designed for a reconstruction of quantum maps from incomplete date can be
modified (see the method of maximum likelihood with an explicitly incorporated condition of the complete positivity) for an
efficient derivation of optimal (given the figure of merit) approximations of un-physical maps.
The derivation of  physical approximations of un-physical maps based on reconstruction schemes as discussed in
the Letter is very natural. This view is supported by the fact that numerical ``reconstructions''
are  compatible with the  analytical results. It should be pointed out that we have chosen two relatively simple examples
(the universal NOT gate that is a non-CP linear map and the nonlinear polarization rotation)
where analytical solutions can be found due to intrinsic symmetries of the considered maps. As soon as the maps do not posses
such symmetries analytical calculations are essentially impossible and the power of our method becomes obvious.

%%%%%%%%%%%%%%%%%%%%%%%%%%%%%%%%%%%%%%%%%%%%%%%%%%%%%%%%%%%%%%%%%%%%%%%%%%%%%
\noindent
%{\em Acknowledgements}\newline
This was work supported in part by  the European
Union projects QUPRODIS and CONQUEST, and
by the Slovak Academy of Sciences via the project CE-PI.

%%%%%%%%%%%%%%%%%%%%%%%%%%%%%%%%%%%%%%%%%%%%%%%%%%%%%%%%%%%%%%%%%%%%%%%%%%%%%
%%%%%%%%%%%%%%%%%%%%%%%%%%%%%%%%%%%%%%%%%%%%%%%%%%%%%%%%%%%%%%%%%%%%%%%%%%%%%

%%%%%%%%%%%%%%%%%%%%%%%%%%%%%%%%%%%%%%%%%%%%%%%%%%%%%%%%%%%%%%%%%%%%%%%%%%%%%
%%%%%%%%%%%%%%%%%%%%%%%%%%%%%%%%%%%%%%%%%%%%%%%%%%%%%%%%%%%%%%%%%%%%%%%%%%%%%
\end{multicols}
\end{document}